# Spin Hall Effect in Collinear Ferromagnets from Spin-Group Symmetry

Qing Zhang[1], Yizhuo Song[1], Jiahao Shentu[1], Jianting Dong[1] and Jia Zhang[1*]

[1]*School of Physics and Wuhan National High Magnetic Field Center,*

*Huazhong University of Science and Technology, 430074 Wuhan, China.*

[*]jiazhang@hust.edu.cn

## Abstract

Magnetic materials support both time-reversal-even (*T*-even) and time-reversal-odd (*T*-odd) spin Hall currents, yet their underlying microscopic origins remain elusive. Here, we elucidate the spin Hall effect (SHE) in collinear ferromagnets by treating spin-orbit coupling (SOC) as a perturbation that breaks spin-group symmetry, thereby revealing how magnetic order activates distinct spin Hall response. To first order in SOC, we identify two dominant *T*-even SHE mechanisms: a magnetization-independent conventional contribution and a magnetization-dependent channel associated with anomalous Hall charge transport. At the same order, the leading *T*-odd magnetic spin Hall effect (MSHE) originates from the exchange interaction between the conventional spin current and the local magnetization. At second order in SOC, we further uncover a distinct *T*-odd planar spin Hall mechanism. Our spin-symmetry analysis is corroborated by first-principles calculations, which reveal a pronounced anisotropic magnetic spin Hall effect whose magnitude can be comparable to the *T*-even spin Hall conductivity (SHC) when the magnetic moment is tilted away from the principal crystallographic axes. These findings clarify the microscopic origins of the SHC in collinear ferromagnets and pave the way for ferromagnet-based spin current sources with versatile properties in spintronic applications.

*Introduction*—The spin Hall effect (SHE) is a central mechanism for spin–charge interconversion and underpins spin–orbit torque (SOT) phenomena in spintronics [1][2][3][4][5]. In nonmagnetic heavy metals, the SHE arises from spin–orbit coupling (SOC), which converts a longitudinal charge current into a transverse spin current under an applied electric field[6][7][8].The pursuit of spin current with higher efficiency and tunability, however, has stimulated growing interest in spin Hall transport in magnetic systems[9][10]. In ferromagnets, spontaneous magnetization gives rise to exchange-split electronic structures in which spin transport is intrinsically coupled to magnetic order[11]. As a result, magnetic systems can generate spin Hall currents with spin polarizations beyond the conventional restrictions, offering additional degrees of freedom for SOT generation and related spintronic device applications. In particular, spin currents with out-of-plane spin polarization in magnetic metals enable field-free switching of perpendicular magnetization[12][13][14][15][16].

The symmetry-imposed spin Hall conductivity (SHC) tensors in magnetic materials are fully constrained by magnetic point-group (MPG) symmetries[17]. While this symmetry-based description is rigorous at the macroscopic level, it provides limited insight into the microscopic origins of the SHE. Although SHCs in ferromagnets have been investigated via first-principles calculations[18][19][20][21][22] and phenomenological analyses[23], their underlying microscopic mechanisms remain poorly understood. As the microscopic origin of the SHE, SOC dictates its magnitude, and angular dependence in ferromagnets. In magnetic systems, SOC interplay with magnetic order enables additional response channels, raising the question of how spin Hall responses evolve with SOC strength. Specifically, it remains unresolved how these spin Hall response maps onto a perturbation expansion in SOC strength, and what microscopic scattering processes govern each individual contribution[24]. A comprehensive understanding of these mechanisms is thus of fundamental importance for the development of next-generation spintronics devices[25][26].

To address this gap, an SOC-order-resolved decomposition of the SHC is critical to establish a microscopic classification of its constituent contributions. Such a framework enables systematic identification of the dominant terms, clarifies their dependence on the magnetization direction, and provides a rigorous basis for the quantitative disentanglement of competing mechanisms within first-principles calculations.

In this work, we develop a microscopic theory of the SHE in collinear ferromagnets by treating SOC as a perturbation that breaks spin-group symmetry, formulated within the spin–orbit vector formalism[26]. Applying this framework to face-centered cubic (fcc) ferromagnets, we identify the leading mechanism underlying both time-reversal-even ($T$-even) and time-reversal-odd ($T$-odd) SHCs and uncover a prominent planar Hall contribution to the magnetic SHC. We quantitatively validate our theoretical predictions by first-principles calculations for representative Ni-based fcc ferromagnets, including Ni, $Ni_{0.8}Fe_{0.2}$ permalloy, and $Ni_{0.5}Pt_{0.5}$ alloy, yielding a unified picture of the SHE in collinear ferromagnets.

*Expansion of SHC on SOC vectors and spin Hall mechanism*—Within the linear response formalism, the electric field-induced spin current can be written as:

$$J_i^k = \sigma_{ij}^k E_j \tag{1}$$

where indices $i$, $j$, and $k$ label the spin current, electric field, and spin polarization directions, respectively, in Cartesian coordinates. By introducing three spin-orbit vectors $\boldsymbol{O}^a$ ($a$=1, 2, 3), the SHC can be expanded in increasing powers of the SOC strength as:

$$\sigma_{ij}^k = \sigma_{ij}^{k(0)} + \sum_{a,l} \alpha_{ij,l}^{k,a} O_l^a + \sum_{ab,lm} \beta_{ij,lm}^{k,ab} O_l^a O_m^b + \cdots\cdots \tag{2}$$

where $\sigma_{ij}^{k(0)}$ denote the non-relativistic SHC. The $\boldsymbol{\alpha}$ and $\boldsymbol{\beta}$ tensors are coefficients proportional to the first and second powers of the SOC strength $\xi$, *i.e.* $\alpha \propto \xi$ and $\beta \propto \xi^2$. The superscripts $a$, $b$..., and subscripts $l$, $m$... of the $\boldsymbol{O}$ vectors index the spin and orbital spaces, respectively. The vectors $\boldsymbol{O}^a$ form a coordinate basis in spin space satisfying $\boldsymbol{O}^1 \times \boldsymbol{O}^2 = \boldsymbol{O}^3$ with $\boldsymbol{O}^3 = \boldsymbol{m} = (m_x, m_y, m_z)$ aligned along the magnetization direction.

This approach also naturally decomposes the SHC tensors into $T$-even and $T$-odd components, which satisfy $\sigma_{ij}^k(\boldsymbol{m}) = \sigma_{ij}^k(-\boldsymbol{m})$ and $\sigma_{ij}^k(\boldsymbol{m}) = -\sigma_{ij}^k(-\boldsymbol{m})$, respectively. Following the spin-group formalism[27][28][29][30], a spin point-group (SPG) operation is denoted as $\{U||R\}$, where $U$ and $R$ act independently in spin space and real space, respectively. The spin-orbit vector then transforms according to:

$$O_l^a = \det(U)\det(R)R_{ll'}U_{aa'}O_{l'}^{a'} \quad (3)$$

where $\det(U)$ and $\det(R)$ are the determinants of the respective symmetry matrices. The symmetry-allowed tensor coefficients $\boldsymbol{\alpha}$ and $\boldsymbol{\beta}$ are fully constrained by their transformations under SPG operations. For clarity, the explicit expressions for the first-order SOC coefficients $\boldsymbol{\alpha}$ and the second-order SOC $\boldsymbol{\beta}$ coefficients, resolved into $T$-odd and $T$-even components, are provided in the Supplementary Note 1.

In this work, we focus on collinear ferromagnets, whose SPG decomposes into the direct product of a collinear spin-only group and a nontrivial spin group. For collinear magnets, the spin-only group consists of elements $\{C_{2n}T||E\}$ and $\{C_{\varphi}||E\}$: $C_{2n}$ denotes a twofold spin rotation about an axis perpendicular to the spin quantum axis, while $C_{\varphi}$ denotes an arbitrary spin rotation by angle $\varphi$ about this same axis. Here, $T$ is the time reversal operation, equivalent to inversion in spin space, and $E$ is the identity operation in real space. The nontrivial spin group is isomorphic to the nonmagnetic crystal point group of the host collinear ferromagnet. For prototypical fcc ferromagnets, this nontrivial group corresponds to the $O_h$ point group, with generators $C_{4z}$, spatial inversion $P$, and $C_{3[111]}$.

We first apply the symmetry constraints of the spin-only group to the $\alpha$ tensor coefficients, yielding nonzero $T$-odd $\alpha$ tensors and $T$-even tensors $\alpha_{ij,l}^{1,1} = \alpha_{ij,l}^{2,2}$, $\alpha_{ij,l}^{3,3}$. Upon further imposing the nonmagnetic point group symmetry, the $T$-odd SHC coefficients at first order in SOC reduce to a single independent parameter $a \propto \xi$. The resulting $T$-odd SHCs *i.e.* magnetic spin Hall conductivity (MSHC) in the crystal coordinate system can be expressed as a function of the magnetization unit vector $\boldsymbol{m}$, and are given by (See Supplementary Note 2 for details):

$$\sigma_{zx}^{y(1)} = 0, \sigma_{zx}^{x(1)} = a^{(1)}m_z, \sigma_{zx}^{z(1)} = -a^{(1)}m_x; \sigma_{zy}^{x(1)} = 0, \sigma_{zy}^{y(1)} = a^{(1)}m_z, \sigma_{zy}^{z(1)} = -a^{(1)}m_y \quad (4)$$

Here the superscript (1) on SHCs denotes the linear (first-order) scaling with SOC strength.

As shown in Fig. 1(a), we focus on the spin current propagating along the $z$

direction driven by an in-plane electric field $\boldsymbol{E}_{xy}=(E_x, E_y, 0)$. The resulting SHC takes the vector form:

$$\boldsymbol{J}_z^{(1),\mathrm{odd}} = a^{(1)}\boldsymbol{m}\times(\boldsymbol{E}_{xy}\times\boldsymbol{z}) \quad (5)$$

Similarly, for fcc collinear ferromagnets, the $T$-even SHC coefficients at first order in SOC are derived via the same symmetry procedure and are characterized by two independent parameters $b$ and $c$, both proportional to $\xi$. The resulting $T$-even SHC components are given by (See Supplementary Note 2 for details):

$$\begin{aligned}&\sigma_{zx}^{x(1)} = -c^{(1)}m_x m_y, \sigma_{zx}^{y(1)} = b^{(1)} - c^{(1)}m_y^2, \sigma_{zx}^{z(1)} = -c^{(1)}m_y m_z;\\&\sigma_{zy}^{x(1)} = -b^{(1)}+c^{(1)}m_x^2, \sigma_{zy}^{y(1)} = c^{(1)}m_x m_y, \sigma_{zy}^{z(1)} = c^{(1)}m_x m_z\end{aligned} \quad (6)$$

The spin current propagating along the $z$ direction, driven by an in-plane electric field, can be expressed in vector form as:

$$\boldsymbol{J}_z^{(1),\mathrm{even}} = b^{(1)}(\boldsymbol{z}\times\boldsymbol{E}_{xy}) + c^{(1)}[\boldsymbol{z}\cdot(\boldsymbol{m}\times\boldsymbol{E}_{xy})]\boldsymbol{m} \quad (7)$$

The first term corresponds to the magnetization-independent $T$-even conventional spin Hall effect (CSHE), sharing the same microscopic origin as the SHE in nonmagnetic heavy metals, where the electric field, spin-current direction, and spin polarization are mutually orthogonal. The second term arises from the anomalous charge Hall effect at first order in SOC, namely the spin anomalous Hall effect (SAHE), with spin polarization collinear with the magnetization.

As illustrated in Fig. 1(b), a spin current with polarization $\boldsymbol{\sigma}$ incident on a magnetic region with magnetization $\boldsymbol{m}$ acquires a transmitted component proportional to $\mathrm{Im}(t_\uparrow t_\downarrow^*)(\boldsymbol{m}\times\boldsymbol{\sigma})$, where $t_\uparrow$ and $t_\downarrow$ are the spin-dependent transmission coefficients. As summarized in Table I, the MSHC at first-order in SOC in eq.(5) can thus be interpreted as arising from the interaction between the $T$-even first-order SOC spin current and the magnetization, namely $\boldsymbol{J}_z^{(1),\mathrm{odd}} \sim \boldsymbol{m}\times\boldsymbol{J}_z^{(1),\mathrm{even}} \sim \boldsymbol{m}\times(\boldsymbol{E}_{xy}\times\boldsymbol{z})$.

Applying analogous spin-group symmetry constraints to the second-order SOC coefficients $\beta$ (See Supplementary Note 3 for details), we derive the $T$-odd spin current propagating along the $z$ direction driven by an in-plane electric field in the following vector form:

$$J_z^{(2),\mathrm{odd}} = a^{(2)} m_z \boldsymbol{E}_{xy} + b^{(2)} (\boldsymbol{m}\cdot\boldsymbol{E}_{xy})\boldsymbol{z} + c^{(2)} m_z (\boldsymbol{m}\cdot\boldsymbol{E}_{xy})\boldsymbol{m} \tag{8}$$

The first two terms are identical in form to those in the first-order SOC MSHC, indicating that the linear-in-$\boldsymbol{m}$ SHC components receive contributions from both first- and second-order SOC. Furthermore, we find that a planar charge Hall effect emerges at second order in SOC: $J_z^{\mathrm{PHE}(2)} \sim m_z(\boldsymbol{m}\cdot\boldsymbol{E})$. Since the charge PHE is a scalar Hall response, combining it with the magnetization vector $\boldsymbol{m}$ lifts it to a spin response[31], thereby defining a planar spin Hall effect (PSHE) with spin polarization parallel to $\boldsymbol{m}$. As summarized in Table I, the corresponding $T$-odd MSHC scales as $m^3$. This establishes that the cubic magnetization dependence of the SHC first emerges at second order in SOC.

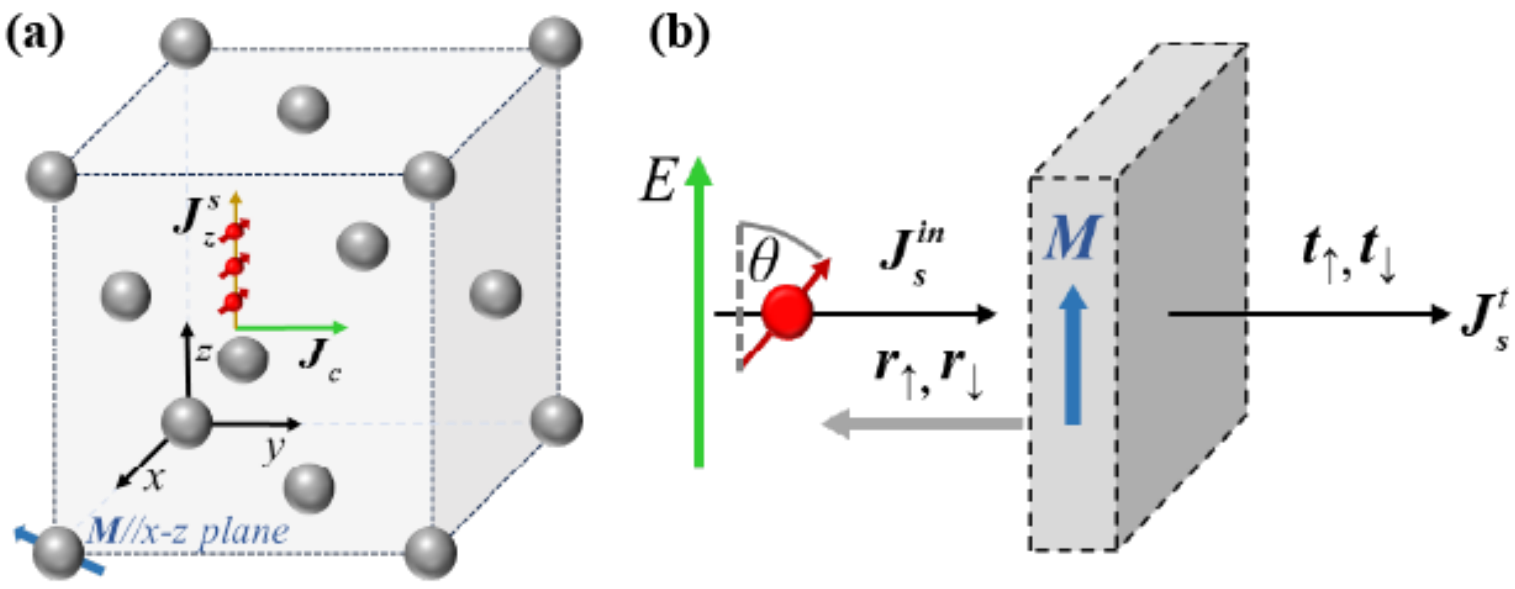

Fig. 1. (a) Generation of a spin Hall current propagating along the $z$ direction by applying an in-plane electric field $\boldsymbol{E}_{xy}$. (b) Schematic of the exchange interaction between the electric field-induced spin Hall current with spin polarization $\boldsymbol{\sigma}$ and the local magnetization $\boldsymbol{m}$. $\theta$ denotes the angle between $\boldsymbol{\sigma}$ and $\boldsymbol{m}$. ($r_\uparrow$, $r_\downarrow$) and ($t_\uparrow$, $t_\downarrow$) are the spin-dependent reflection and transmission coefficients, respectively.

The second-order SOC contribution to the $T$-even spin current propagating along the $z$ direction and driven by an in-plane electric field reads:

$$J_z^{(2),\mathrm{even}} = d^{(2)}[\boldsymbol{z}\cdot(\boldsymbol{m}\times\boldsymbol{E}_{xy})]\boldsymbol{m} + e^{(2)}(\boldsymbol{m}\cdot\boldsymbol{E}_{xy})(\boldsymbol{z}\times\boldsymbol{m}) + f^{(2)} m_z (\boldsymbol{E}_{xy}\times\boldsymbol{m}) \tag{9}$$

The first term represents the second-order SOC contribution to the anomalous charge Hall transport, whereas the last two terms take the form of $J_z^{(2),\mathrm{odd}}\times\boldsymbol{m}$. This indicates that the $T$-even SHC dependence on the square term $m^2$ contains both first- and second-order SOC contributions.

Evaluating higher-order SOC contributions to the SHCs via a full expansion of SOC vectors is computationally cumbersome and lies beyond the scope of this work.

Nevertheless, the nonvanishing third-order SOC coefficients for fcc ferromagnets , as constrained by SPG symmetry, are provided in Supplementary Note 4. In some cases, as discussed later, third- and even higher-order SOC contributions to the SHCs may be sizable and cannot be neglected.

TABLE I. Origins of dominant magnetization-induced spin Hall mechanisms in collinear ferromagnets. MSHE: magnetic spin Hall effect; CSHE: conventional spin Hall effect; PHE: planar Hall effect; AHE: anomalous Hall effect.

| **Type of SHE** | **Microscopic Mechanism** | |
|---|---|---|
| *T*-odd SHE (MSHE) | $\boldsymbol{J}_z^{(1),\mathrm{odd}} \propto \boldsymbol{m}\times(\boldsymbol{E}_{xy}\times\boldsymbol{z})$ | $\boldsymbol{J}_z^{\mathrm{CSHE}(1)} \propto (\boldsymbol{z}\times\boldsymbol{E})$ |
| | $\boldsymbol{J}_z^{(2),\mathrm{odd}} \propto m_z(\boldsymbol{m}\cdot\boldsymbol{E}_{xy})\boldsymbol{m}$ | $J_z^{\mathrm{PHE}(2)} \propto m_z(\boldsymbol{m}\cdot\boldsymbol{E})$ |
| *T*-even SHE | $\boldsymbol{J}_z^{(1),\mathrm{even}} \propto [\boldsymbol{z}\cdot(\boldsymbol{m}\times\boldsymbol{E}_{xy})]\boldsymbol{m}$ | $J_z^{\mathrm{AHE}(1)} \propto [\boldsymbol{z}\cdot(\boldsymbol{m}\times\boldsymbol{E})]$ |
| | $\boldsymbol{J}_z^{(2),\mathrm{even}} \propto [\boldsymbol{z}\cdot(\boldsymbol{m}\times\boldsymbol{E}_{xy})]\boldsymbol{m}$ | $J_z^{\mathrm{AHE}(2)} \propto [\boldsymbol{z}\cdot(\boldsymbol{m}\times\boldsymbol{E})]$ |

*The SOC strength and* ***m*** *dependence of SHC by first-principles calculations*—We first examine the SOC strength dependence of the SHCs for a fixed magnetization orientation and corresponding MPG. With the magnetic moment ***m*** aligned along the *x* axis, there are ten MPG symmetry-imposed nonzero SHC components associated with five independent tensors. These include contributions up to second order in SOC: the *T*-odd MSHC, $\sigma_{yx}^{y}=\sigma_{zx}^{z}=-a^{(1)}+b^{(2)},\sigma_{xy}^{y}=\sigma_{xz}^{z}=a^{(1)}+a^{(2)}$ , the dominant magnetization-independent conventional *T*-even SHC: $\sigma_{zx}^{y}=-\sigma_{yx}^{z}=b^{(1)}+e^{(2)}$ , $\sigma_{xy}^{z}=-\sigma_{xz}^{y}=b^{(1)}-f^{(2)}$, and the *T*-even SHC arising from the interplay between the conventional SHE and anomalous charge Hall contribution $\sigma_{zy}^{x}=-\sigma_{yz}^{x}=-c^{(1)}+d^{(2)}$. From the SOC expansion, the two MSHC components $\sigma_{zx}^{z}$ and $\sigma_{xz}^{z}$ exhibit equal magnitude and opposite sign at first order in SOC, but acquire distinct second-order SOC corrections. In contrast, the two *T*-even SHC components $\sigma_{yx}^{z}$ and $\sigma_{zx}^{y}$ share the same magnetization-independent conventional SHC and are therefore expected to have comparable magnitude.

We can then evaluate SHCs on the increasing orders of relative SOC strength $\xi/\xi_0$ as follows:

$$\sigma_{ij}^{k} = \lambda_{ij}^{k(1)}(\xi/\xi_0) + \lambda_{ij}^{k(2)}(\xi/\xi_0)^2 + \lambda_{ij}^{k(3)}(\xi/\xi_0)^3 ... = \sigma_{ij}^{k(1)} + \sigma_{ij}^{k(2)} + \sigma_{ij}^{k(3)} + ... \quad (10)$$

where $\lambda_{ij}^{k(1),(2),(3)...}$ are the coefficients, $\xi_0$ is the original SOC strength and $\xi$ is the scaled SOC which can be tuned by adjusting the speed of light in the first-principles calculations. This method enables a validation between the above spin-group symmetry analysis and calculation results. Fitting the computed SHCs as a function of the SOC scaling factor $\xi/\xi_0$ allows us to extract the SOC-order dependence of each SHC component and identify the leading contributions in different materials.

The SOC-scaling dependence of the calculated SHCs in $Ni_{0.8}Fe_{0.2}$ is shown in Fig. 2 (The results for Ni and NiPt are shown in Fig. S1 and S2 in the Supplementary Material). All SHC components vanish in the absence of SOC and increase monotonically with SOC strength, confirming their SOC-driven origin in collinear ferromagnets. In $Ni_{0.8}Fe_{0.2}$, the MSHCs are smaller in magnitude than the $T$-even SHCs, which are dominated by the CSHE. For magnetization along $x$, all SHC components of the fcc ferromagnets are adequately described by SOC expansion up to third order. The fitted coefficients, summarized in Table S1 of the Supplementary Material, show that the relative contributions of the first-, second-, and third-order SOC terms are strongly material dependent.

For both $Ni_{0.8}Fe_{0.2}$ and $Ni_{0.5}Pt_{0.5}$, the SHCs can be well fitted up to second-order SOC. As depicted in Fig.2(c) and (d) the representative $T$-even SHC $\sigma_{yx}^{z}$ and MSHC $\sigma_{zx}^{z}$ of $Ni_{0.8}Fe_{0.2}$ are dominated by the first and second order contributions, the third term is negligible. By contrast, for fcc Ni inclusion of the third-order SOC contribution is indispensable for accurately reproducing the calculated SHCs.

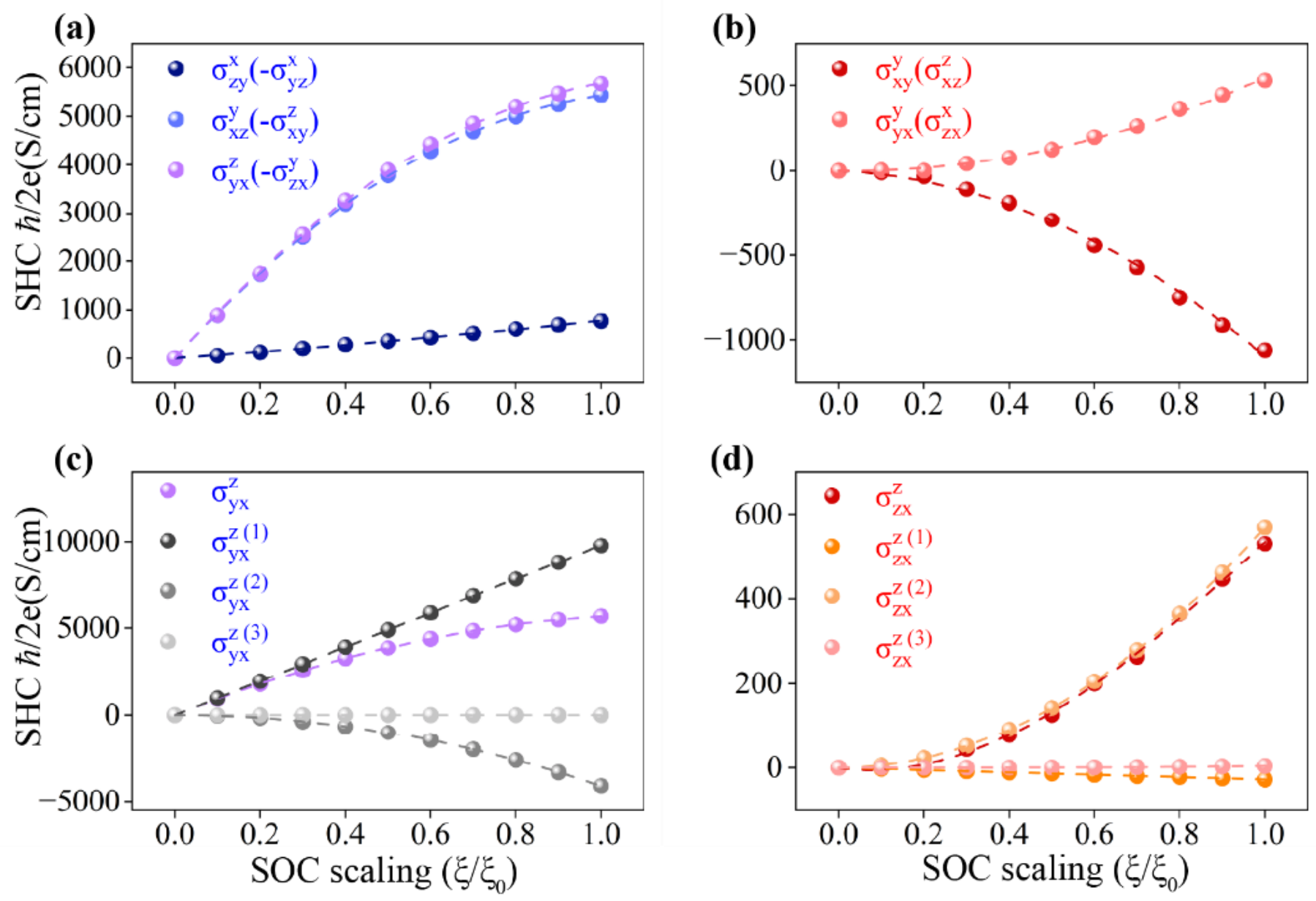

Fig. 2. (a) $T$-even and (b) $T$-odd SHCs as a function of the SOC scaling factor $\xi/\xi_0$ for fcc $Ni_{0.8}Fe_{0.2}$ with magnetization aligned along the $x$ axis. Symbols denote first-principles results, and dashed lines correspond to SOC expansions up to third-order. (c) and (d) show the SOC-order decomposition of $\sigma_{yx}^{z}$ and $\sigma_{zx}^{z}$, respectively.

*The **m** dependence of SHC by first-principles calculations*—Rotating the magnetization within the $xz$ plane of the conventional fcc cell yields the following SHC components up to second order in SOC:

$$\sigma_{zx}^{x} = a^{(1)}m_z + a^{(2)}m_z + c^{(2)}m_x^2 m_z,$$

$$\sigma_{zx}^{y} = b^{(1)} + e^{(2)}m_x^2 - f^{(2)}m_z^2,$$

$$\sigma_{zx}^{z} = -a^{(1)}m_x + b^{(2)}m_x + c^{(2)}m_x m_z^2,$$

$$\sigma_{zy}^{x} = -b^{(1)} + c^{(1)}m_x^2 + f^{(2)}m_z^2 + d^{(2)}m_x^2.$$

With $\theta$ as the polar angle of $\boldsymbol{m}$ from the $z$ axis ($m_x$=sin$\theta$ $m_z$=cos$\theta$), the angular dependence of the SHCs can be derived analytically. As shown in Fig.3(a)(b), for $Ni_{0.8}Fe_{0.2}$ (the results for Ni can be found in Supplementary Fig. S3), the $T$-even SHC $\sigma_{zx}^{y}$ exhibits only weak angular dependence, reflecting its dominant CSHE origin. In contrast, the $T$-even SHC $\sigma_{zy}^{x}$ containing a sizable anomalous Hall contribution likewise shows obvious angular dependence. The MSHCs $\sigma_{zx}^{x}$ , $\sigma_{zx}^{z}$ shown in Fig.3(c)(d) contain a substantial planar Hall contribution, exhibiting strong angular dependence. Notably, when the magnetization is rotated away from the principal crystal axes, the magnitude of the MSHCs becomes comparable to that of the *T*-even SHCs. One MSHC component $\sigma_{zx}^{x}$ reaches $-4575\,\hbar/2e$ (S/cm) at $\theta$≈55°, whereas the other $\sigma_{zx}^{z}$ reaches $-3650\,\hbar/2e$ S/cm at $\theta$≈35°. Specially, the SHC $\sigma_{zx}^{x}$ with *z*-polarization is of great importance and useful for field-free switching of perpendicular magnetization[21].

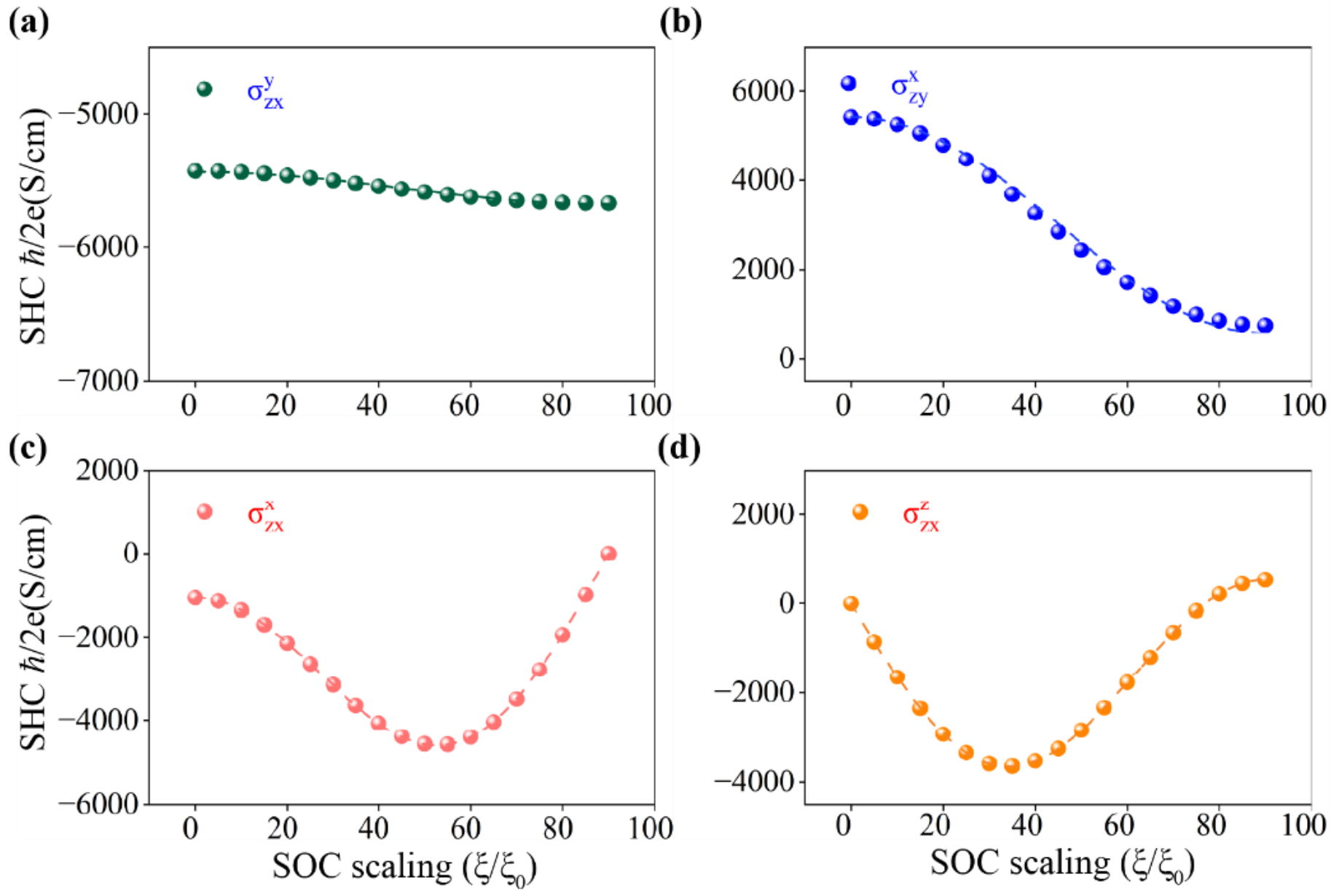

Fig. 3. SHCs of $Ni_{0.8}Fe_{0.2}$ as a function of the angle $\theta$ between the magnetic moment $\boldsymbol{m}$ and the $z$ axis of the conventional fcc unit cell, with $\boldsymbol{m}$ constrained to the $xz$ plane. Symbols represent the first-principles calculation results, and dashed lines show fitting curves corresponding to the analytical expressions derived from the spin-symmetry allowed second-order SOC expansion.

Table II summarizes the key fitting coefficients describing the angular dependence of the SHCs in fcc $Ni_{0.8}Fe_{0.2}$ and Ni up to second order in SOC. In both materials, the coefficients for *T*-even conventional SHE and spin anomalous Hall have comparable magnitudes. The planar Hall coefficients associated with *T*-odd MSHCs, though of second order in SOC, are nonetheless substantial, which reflects a significant planar

Hall contribution to the total MSHC. It is also noteworthy that while this second-order expansion captures a substantial part of the SHCs, it does not fully account for the angular dependence of all SHC tensor elements. As shown in Supplementary Fig. S4, the angular dependence of two other SHC components, $\sigma_{zy}^{y}$ and $\sigma_{zy}^{y}$, cannot be fully reproduced within a second-order SOC expansion. Higher-order terms are therefore required to describe their full anisotropic SHC response.

TABLE II. The fitting coefficients to calculated angular-dependent SHCs with first- and second-order SOC contributions according to different mechanisms for fcc-$Ni_{0.8}Fe_{0.2}$ and Ni and $Ni_{0.5}Pt_{0.5}$. For clarity, this Table lists the PSHE separately from the other magnetic spin Hall effect (MSHE) contributions according to its distinct microscopic mechanism, although it is formally part of the MSHE.

| | *T*-odd | | | *T*-even | |
|---|---|---|---|---|---|
| | | ***m***-dependent | | ***m***-independent | ***m***-dependent |
| | MSHC | MSHC | PSHC | CSHC | SAHC |
| $Ni_{0.8}Fe_{0.2}$ | $a^{(1)}= -1152$ | $a^{(2)}=91$, $b^{(2)}=-626$ | $c^{(2)}=-10322$ | $b^{(1)}=-2405$ | $c^{(1)}+d^{(2)}=-1810$ |
| Ni | $a^{(1)}=-1493$ | $a^{(2)}= -633$, $b^{(2)}= -437$ | $c^{(2)}=-7570$ | $b^{(1)}=-2800$ | $c^{(1)}+d^{(2)}=-1408$ |

*Summary*—In this work, we develop a microscopic framework for the SHE in collinear ferromagnets based on spin-group symmetry and a perturbation expansion of SOC. By decomposing the SHC tensors into contributions from successive SOC orders, this framework elucidates the symmetry origins and underlying microscopic mechanisms of each spin Hall response. We reveal that the first-order SOC contribution to the magnetization-induced SHC (MSHC) and identify a prominent plan Hall mechanism emerging at second-order SOC.

We validate this expansion scheme via combined symmetry analysis and first-principles calculations on fcc ferromagnets. We identify a pronounced angular dependence of the SHCs, which is quantitatively corroborated by first-principles calculations. Notably, MSHCs can reach the same order of magnitude as conventional SHCs when the magnetization is not aligned along the principle crystal axes. For collinear ferromagnets, the dominant angular dependence of the SHC is well captured by SOC contributions up to second order.

A key strength of this approach is that the spin-group symmetry explicitly links magnetic order, SOC, and physical observables, enabling quantitative analysis beyond conventional symmetry-based classifications. The framework can be readily generalized to collinear antiferromagnets by replacing the magnetic order parameter $\boldsymbol{m}$ with the Néel vector $\boldsymbol{l}$[36], and extended to other SOC-driven spin-transport phenomena in magnetic materials, including the Edelstein effect[37] and spin-orbit torques[3].

*Acknowledgement*—This work was supported by the National Natural Science Foundation of China (Grants No. T2394475, No. T2394470), the National Key Research and Development Program of China (2024YFA1611200), and the open research fund of Beijing National Laboratory for Condensed Matter Physics (Grant No. 2023BNLCMPKF010).